\documentclass[journal]{IEEEtran}

\usepackage{amsmath}
\usepackage{amssymb}
\usepackage{bm}

\usepackage{graphicx}

\usepackage{float}

\usepackage{xcolor,stfloats}
\usepackage{lipsum}

\usepackage{algorithm, algpseudocode}

\usepackage{ifpdf}

\usepackage{cite}
\usepackage{hyperref}

\ifCLASSINFOpdf
\else
\fi
\usepackage{amsmath}
\usepackage{url}

\begin{document}


\title{Semantic Communication for the Internet of Space: New Architecture, Challenges, and Future Vision}

%

\author{Hanlin~Cai,~\IEEEmembership{Student~Member,~IEEE,} 
        Houtianfu~Wang,~\IEEEmembership{Student~Member,~IEEE,} 
        Haofan~Dong,~\IEEEmembership{Student~Member,~IEEE,} 
        and~Ozgur~B.~Akan,~\IEEEmembership{Fellow,~IEEE}


\thanks{The authors are with the Internet of Everything (IoE) Group, Electrical Engineering Division, Department of Engineering, University of Cambridge, CB3 0FA Cambridge, U.K. (e-mail: hc663@cam.ac.uk, hw680@cam.ac.uk, hd489@cam.ac.uk, oba21@cam.ac.uk).}
\thanks{O. B. Akan is also with the Center for neXt-Generation Communications (CXC), Department of Electrical and Electronics Engineering, Koç University, 34450 Istanbul, Turkey (e-mail: akan@ku.edu.tr).}

}

%
%


\markboth{SUBMITTED TO IEEE COMMUNICATIONS STANDARDS MAGAZINE}%
{}

%




\maketitle


\begin{abstract}

The expansion of sixth-generation (6G) wireless networks into space introduces technical challenges that conventional bit-oriented communication approaches cannot efficiently address, including intermittent connectivity, severe latency, limited bandwidth, and constrained onboard resources. To overcome these limitations, semantic communication has emerged as a transformative paradigm, shifting the communication focus from transmitting raw data to delivering context-aware, mission-relevant information. In this article, we propose a semantic communication architecture explicitly tailored for the 6G Internet of Space (IoS), integrating multi-modal semantic processing, AI-driven semantic encoding and decoding, and adaptive transmission mechanisms optimized for space environments. The effectiveness of our proposed framework is demonstrated through a representative deep-space scenario involving semantic-based monitoring of Mars dust storms. Finally, we outline open research challenges and discuss future directions toward realizing practical semantic-enabled IoS systems.

\end{abstract}

\begin{IEEEkeywords}
Semantic Communication, Internet of Space.
\end{IEEEkeywords}

\IEEEpeerreviewmaketitle


\section{Introduction}

Sixth-generation (6G) wireless networks promise to extend connectivity beyond traditional terrestrial boundaries, giving rise to the \emph{Internet of Space} (IoS), an integrated communication fabric seamlessly interconnecting satellites, spacecraft, airborne platforms, and terrestrial infrastructures. IoS is envisioned to support a diverse range of mission-critical applications, including global broadband connectivity, Earth observation, real-time environmental monitoring, and autonomous space exploration, each demanding stringent performance metrics such as ultra-low latency, ultra-high reliability, and massive device connectivity~\cite{jiao2021massive}. However, conventional terrestrial-oriented communication frameworks encounter significant limitations in addressing the unique challenges posed by space environments, such as intermittent connectivity due to orbital dynamics, severe bandwidth and energy constraints, prolonged propagation delays in interplanetary links, and the inherent complexity of multi-modal data types~\cite{meng2024semantics}.

Recently, semantic communication has emerged as a transformative paradigm, fundamentally shifting the communication emphasis from transmitting raw bits to conveying meaningful, task-oriented information. By intelligently identifying, encoding, and transmitting only the most relevant semantic content, this approach significantly reduces redundant information and enhances resource efficiency, making it particularly attractive for resource-constrained space missions~\cite{sagduyu2024will}. Recent studies have demonstrated the potential of semantic encoding techniques, particularly those leveraging deep learning, to substantially improve spectrum utilization and reliability, achieving superior task performance with substantially reduced data transmission requirements~\cite{zhang2024semantically, wang2023ca_deepsc}.

While initial efforts to incorporate semantic communication principles into IoS show promise, critical research gaps remain unaddressed. Existing studies predominantly focus on terrestrial and near-Earth scenarios, lacking dedicated semantic communication frameworks tailored explicitly for deep-space or interplanetary environments characterized by extreme propagation delays, high error rates, and intermittent links~\cite{yang2023environment, xu2024federated}. Moreover, current approaches largely neglect comprehensive considerations of standardized semantic interoperability across heterogeneous IoS platforms and fail to adequately address multi-modal semantic data fusion from diverse sources such as hyperspectral imaging, telemetry, and sensor time-series data \cite{wei2024iris}. Although recent research highlights the potential of integrated sensing and communication (ISAC) to acquire multimodal data through joint waveform design and unify perception-communication functionalities~\cite{dong2024martian}, a systematic integration of ISAC with semantic-driven IoS frameworks remains largely unexplored.

To bridge these critical gaps, this article introduces a semantic communication architecture explicitly designed for the 6G Internet of Space (IoS). Unlike existing terrestrial-focused semantic frameworks, our approach systematically addresses the unique constraints of space environments, including intermittent connectivity, significant propagation delays, limited bandwidth, and strict onboard resource limitations. By emphasizing onboard semantic extraction, encoding, and adaptive transmission of mission-specific information, the proposed architecture enhances efficiency and reliability in IoS communications. The key contributions of this article are summarized as follows:

\begin{itemize}
    \item To the best of our knowledge, this is the first semantic communication architecture specifically designed for IoS environments. We propose a three-layer framework, comprising a Data Layer for multi-modal feature extraction, a Transport Layer for semantic encoding and transmission, and an Application Layer for mission-level interpretation and decision-making, targeted at addressing the unique constraints of space communications.

    \item We pioneer the synergistic ISAC with terahertz (THz) links, establishing an intelligent co-design paradigm that dynamically aligns multi-modal sensing fidelity with semantic entropy demands, thereby achieving simultaneous enhancement of spectral efficiency and task-aware robustness in harsh space environments.
    


    \item We validate the proposed architecture through a representative deep-space scenario involving semantic-based monitoring of Mars dust storms, demonstrating substantial improvements in energy efficiency, transmission reliability, and task-oriented decision support.
\end{itemize}




The rest of this article is organized as follows. The requirements and enabling technologies for semantic-enabled IoS communication are first outlined. The proposed semantic IoS architecture is then presented, highlighting its multi-modal processing, semantic encoding, and adaptive transmission mechanisms. Representative deep-space scenarios are provided to demonstrate the effectiveness of the proposed architecture. Potential challenges and future research directions related to semantic IoS communication are subsequently discussed. Finally, conclusions are drawn in the last section.


\begin{figure*}[!t]
\centering
\includegraphics[width=1\linewidth]{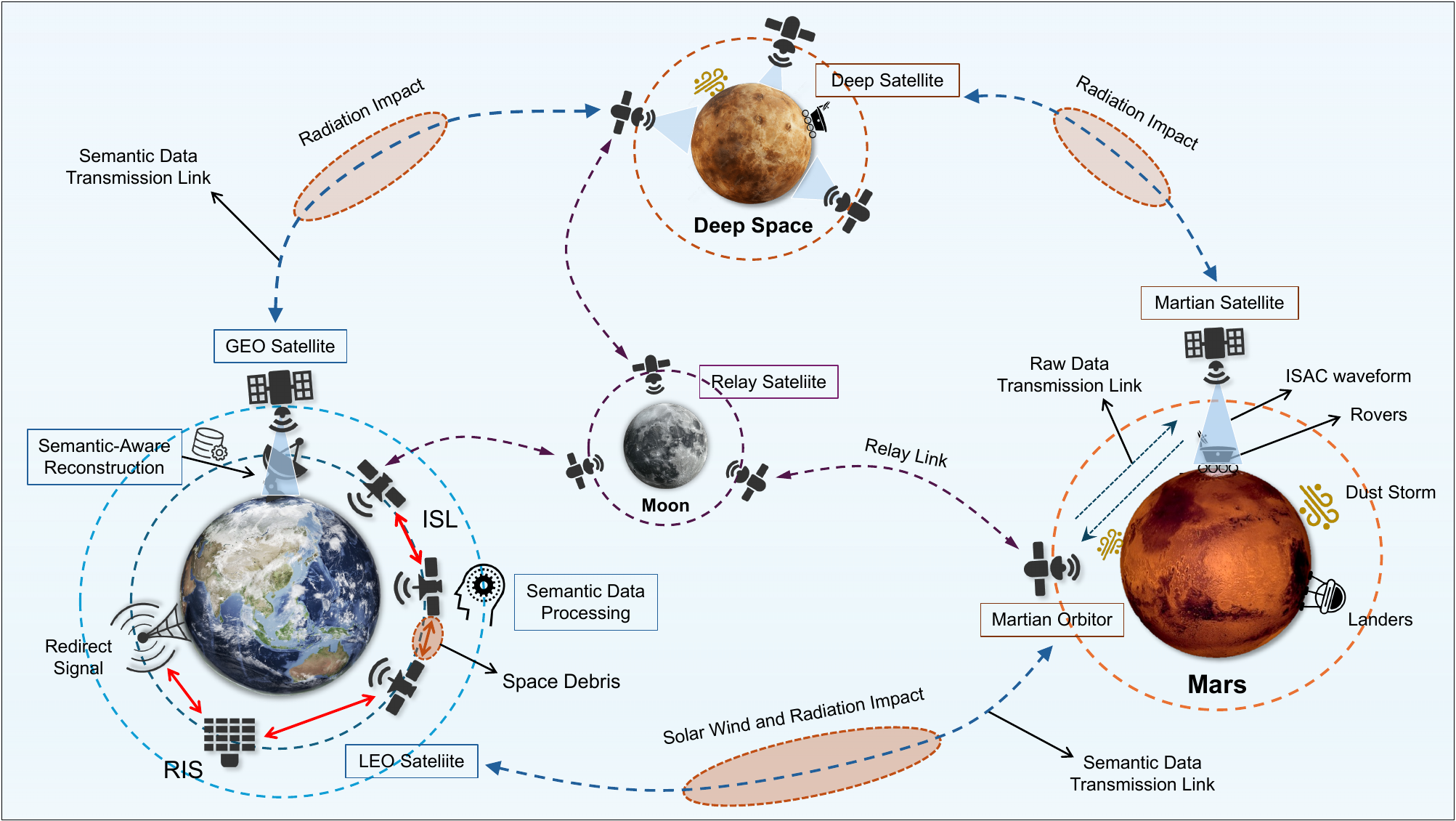}

\caption{Internet of Space semantic communication architecture showing Earth-Mars-Deep space communications as a representative example within the IoS framework. The system illustrates the contrast between traditional raw data transmission and semantic data transmission links. Key components include orbital satellites, relay systems, and surface elements operating under challenging environmental conditions. The semantic encoding and decoding processes demonstrated here can be extended to various solar and outer solar system missions, enabling effective communications despite significant distances, radiation effects, and resource limitations characteristic of space exploration.}
\label{fig:ios_semantic}
\end{figure*}

\section{6G Internet of Space: Challenges and Requirements}

Internet of Space (IoS) represents a significant extension of wireless communications beyond terrestrial boundaries, creating unique challenges that demand innovative approaches in the 6G era. This section examines the distinctive challenges of space communications, outlines essential requirements for next-generation IoS, and demonstrates how semantic communication serves as a transformative enabler for near-Earth to deep space applications.

\subsection{IoS Communication Challenges}

As illustrated in Fig. \ref{fig:ios_semantic}, space communications operate in environments fundamentally different from terrestrial networks. Near-Earth orbital networks face extreme dynamics, with LEO satellites typically moving at approximately 7.8 km/s, creating rapidly changing network topologies and significant Doppler effects. Meanwhile, interplanetary links experience extraordinary path losses (exceeding 200dB) and propagation delays ranging from 4-25 minutes for Earth-Mars communications – conditions where traditional Shannon-based approaches become inefficient or impractical \cite{al2022survey}.

Deep space communications further extend these challenges. For missions to outer planets within our solar system, round-trip light times can reach several hours, with severely constrained power budgets and limited data rates. Such scenarios represent significant technical challenges, where efficient information transfer becomes increasingly critical. Scientific missions generate substantial data volumes, including imagery, spectroscopy, and various sensor measurements, yet can transmit only a fraction back to Earth, creating a fundamental information bottleneck that conventional communication paradigms struggle to address.

The physical space environment introduces additional complexities. Signals traversing the ionosphere (60-1000 km altitude) encounter frequency-dependent refraction and scintillation, particularly affecting sub-3 GHz transmissions \cite{ye2025dancing}. Deep space links face periodic solar plasma interference, cosmic radiation, and signal degradation across vast distances, while planetary and lunar surface networks must contend with harsh local conditions including dust storms, extreme temperature variations, and challenging terrain.

Space platforms operate under severe resource constraints, with strict limitations on power generation, computing capabilities, thermal management, and hardware redundancy. These limitations necessitate fundamentally rethinking how information is processed and transmitted across space, particularly for long-duration missions where communication windows may be brief, unpredictable, or separated by extended periods without contact.

\subsection{Essential Requirements for IoS Communication}

To effectively address the unique communication challenges inherent to the Internet of Space (IoS), next-generation space communication systems must meet several critical requirements that surpass conventional terrestrial approaches. These essential requirements can be summarized into four dimensions:


\textbf{Prioritized Data Management:}  
Due to limited transmission windows and bandwidth constraints in space, IoS systems must differentiate mission-critical data from routine transmissions. Critical information, such as emergency telemetry, vital scientific discoveries, or operational alerts, should be prioritized over standard measurements, ensuring timely and efficient utilization of available communication resources.

\textbf{Adaptive Temporal-Spatial Operation:}  
Unlike terrestrial networks, IoS networks must dynamically adapt their configurations in response to rapidly changing environmental conditions, mission phases, and communication opportunities. This adaptability becomes especially critical during deep-space missions, where spacecraft encounter diverse operational contexts over extended timeframes. Consequently, delay-tolerant networking (DTN) protocols must integrate semantic awareness to intelligently manage data priority based on content relevance and contextual urgency, rather than solely relying on traditional metrics such as arrival time or data format.

\textbf{Autonomous Edge Intelligence:}  
Considering the extensive signal propagation delays experienced in deep-space missions—often ranging from minutes to hours—real-time Earth-based control becomes impractical. Therefore, IoS communication architectures must enable autonomous edge intelligence, empowering space assets to independently evaluate and determine data relevance. Such autonomy involves local semantic processing, real-time decision-making regarding data transmission priorities, and onboard operational adjustments without continuous Earth intervention.

\textbf{Optimized Resource Utilization:}  
Given stringent power and spectrum constraints, IoS communication systems must maximize information value while minimizing energy and bandwidth consumption. Particularly in deep-space scenarios, where spacecraft operate with severely limited energy resources and encounter considerable propagation losses, novel communication techniques are required to deliver maximal semantic value using minimal resources. These solutions must ensure sustainable and reliable information exchange throughout mission lifetimes.


\subsection{Semantic Communication as Enabler}

\begin{figure}
    \centering
    \includegraphics[width=1\linewidth]{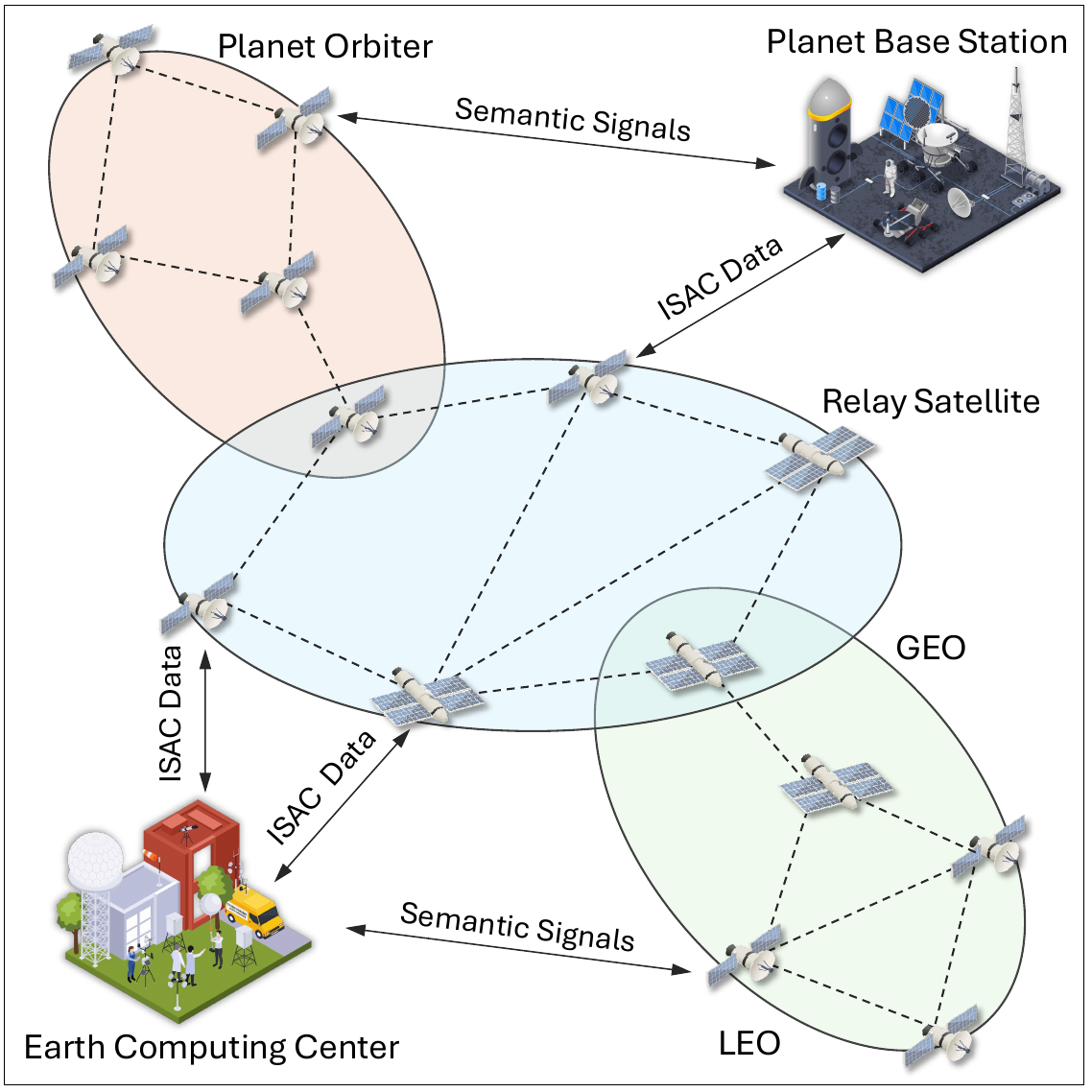}
    \caption{Hierarchical satellite network architecture for semantic-enabled IoS.
    }
    \label{fig:satellite}
\end{figure}

Semantic communication emerges as a key enabler for IoS, working alongside advanced coding, modulation, and networking technologies to address these fundamental challenges through three primary mechanisms:

The environment-aware semantic processing mechanism extracts and prioritizes task-relevant information from raw data, reducing transmission volume while preserving mission-critical content. This capability enables deep space probes to identify and prioritize significant observations for Earth notification, while filtering routine data that conforms to expected parameters. Such processing can substantially reduce communication requirements for bandwidth-constrained links spanning Mars to outer solar system missions, while ensuring important discoveries reach Earth despite distance constraints.

Building upon this foundation, adaptive semantic encoding leverages contextual understanding to optimize resource allocation across diverse space segments. As shown in Fig.~\ref{fig:ios_semantic}, this approach enables efficient communication even in challenging interplanetary scenarios affected by solar wind and radiation. The figure demonstrates Earth-Mars-Deep space communications as a representative case, with the supporting relay infrastructure detailed in Fig.~\ref{fig:satellite} illustrating tiered node coordination across orbital regimes. The principles apply to broader deep space missions—semantic encoding can compress information to essential meanings, achieving significantly improved data efficiency while preserving scientific value.

Complementing these capabilities, distributed intelligence frameworks enable collaborative operations through efficient model parameter sharing among space assets. The hierarchical relay architecture in Fig.~\ref{fig:satellite} exemplifies how orbital layers synergistically implement these frameworks, preserving bandwidth while enhancing system capabilities. This approach creates a semantic knowledge network spanning from Earth orbit to deep space missions. The architecture maintains resilient operations that persist during extended communication outages, allowing distant spacecraft to operate with increased autonomy when direct Earth communication is unavailable.



\section{Semantic Empowered IoS: New Architecture and Standardization}

\begin{figure*}[!t]
    \centering
    \includegraphics[width=1\linewidth]{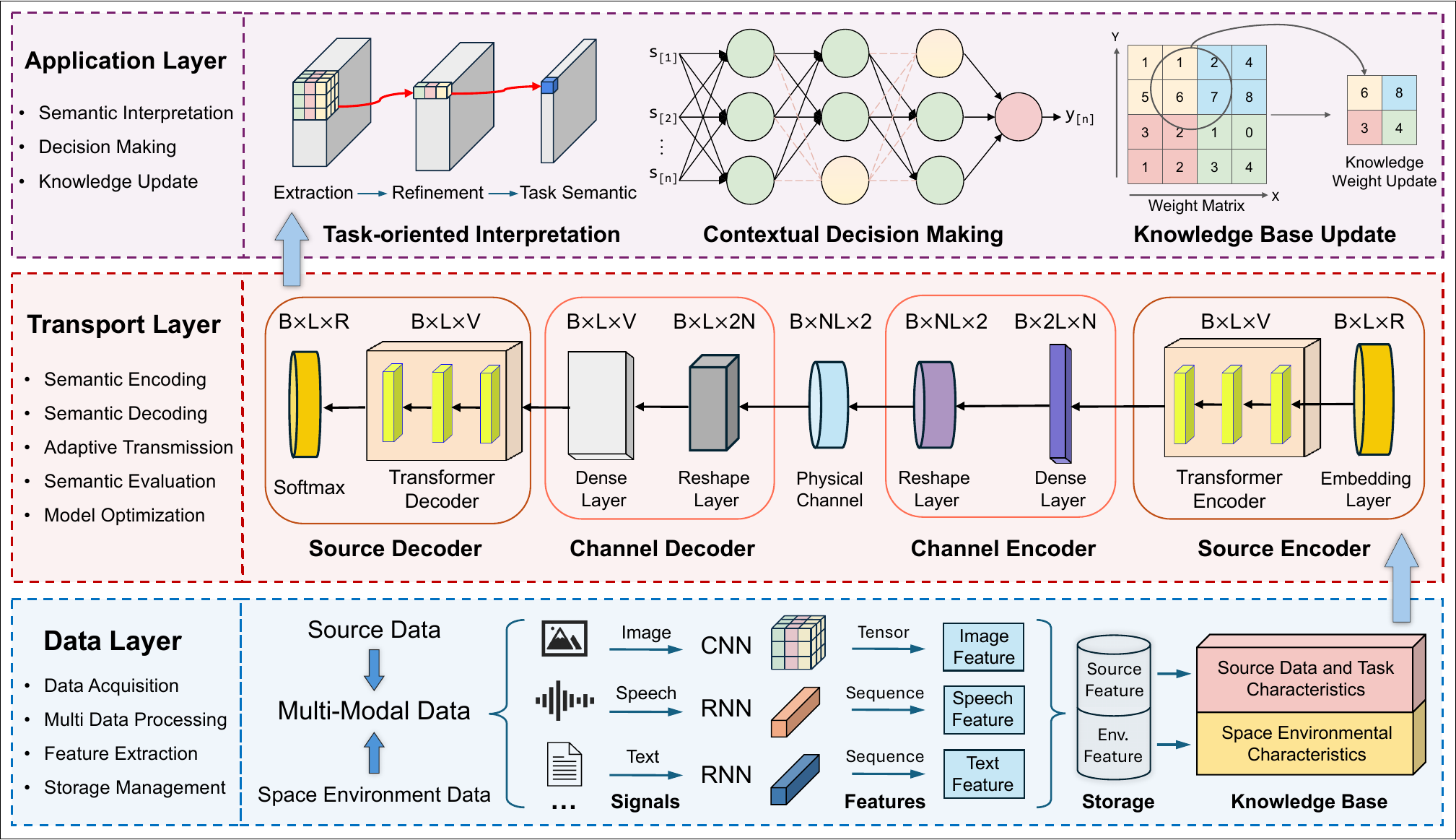}
    \caption{Semantic Empowered IoS: A layered architecture illustrating multi-modal data processing, semantic encoding/decoding, transmission mechanisms, and the integration of standardization for enhanced semantic communication in space.}
    \label{fig:architecture}
\end{figure*}

The integration of semantic communication into the Internet of Space (IoS) requires a structured and scalable architecture capable of supporting multi-modal data acquisition, adaptive information transmission, and mission-driven decision-making. In contrast to conventional space communication systems, which rely heavily on syntactic-level data exchange, the proposed architecture employs semantic intelligence to enhance operational efficiency, resilience, and interoperability across diverse space infrastructures. By organizing the system into distinct functional layers, the architecture enables real-time adaptability, supports legacy compatibility, and optimizes task-specific performance—attributes essential for intelligent and scalable deep-space missions. Fig.~\ref{fig:architecture} presents the proposed three-layer semantic architecture, which addresses these requirements through the following components:

\begin{itemize}
    \item \textbf{Data Layer:}  
    Collects heterogeneous sensor inputs, including imaging, telemetry, and spectral data, and performs on-board feature extraction and cross-modal fusion using lightweight neural models. Adaptive caching ensures retention of mission-critical information during connectivity disruptions.

    \item \textbf{Transport Layer:}  
    Executes task-driven semantic encoding and channel-adaptive transmission with hybrid error correction. Standardized metadata enables flexible delivery while maintaining interoperability and semantic fidelity.

    \item \textbf{Application Layer:}  
    Interprets semantic data through domain-specific knowledge bases and supports real-time, context-aware decision-making. Continuous knowledge refinement ensures autonomous adaptation to dynamic mission environments.
\end{itemize}

Designed with CCSDS/ITU-compliant interfaces and unified metadata structures, such as task identifiers and spatiotemporal tags, the architecture effectively decouples sensing, transmission, and decision-making while enabling both backward compatibility and forward scalability. The following subsections detail the core functions of each layer and demonstrate how the architecture integrates adaptive acquisition, semantic-aware delivery, and evolving knowledge to meet the operational demands and interoperability requirements of future space communication systems.


\subsection{Data Layer}

The Data Layer forms the foundational infrastructure for raw data collection and pre-processing in the IoS semantic communication framework, orchestrating a four-stage pipeline to transform heterogeneous inputs into task-ready semantic features.

Multi Modal Data Acquisition is achieved through Integrated Sensing and Communication (ISAC) technology, which synchronizes heterogeneous sensor systems, including imaging devices (e.g., optical/thermal cameras), telemetry units, and spectral analyzers, to capture diverse data types such as images, text, telemetry signals, and spectral measurements. To enhance this acquisition process, emerging reconfigurable intelligent surfaces (RIS) provide physical-layer enhancements for robust data collection. Mounted on orbital platforms, these metamaterial arrays dynamically reshape wireless propagation characteristics through real-time beamforming. During lunar occultation periods when direct Earth communication is blocked, RIS-equipped relay satellites can maintain data links between surface probes and orbital assets by creating synthetic reflection paths, a capability critical for hyperspectral data transmission requiring preserved semantic features.

Building on the acquired data, Feature Extraction and Fusion employ lightweight neural networks (CNNs, RNNs, or Transformers) deployed directly on satellites or spatial nodes. These models perform localized processing to extract spatial-temporal features from raw data, such as identifying geological structures with CNNs or analyzing telemetry trends with RNNs. Cross-modal fusion mechanisms further combine features from different data streams (e.g., aligning image regions with spectral signatures) to generate unified semantic representations.

Following feature fusion, Data Pre-processing and Filtering ensures data quality through noise reduction, anomaly detection, and redundancy removal. Telemetry data is filtered in real-time to transmit only critical anomalies like orbital deviations, while imaging data is cleansed of sensor noise and compressed by retaining mission-critical areas such as space debris or thermal anomalies.

To address intermittent connectivity challenges introduced by pre-processing, Data Storage and Management implements adaptive buffering and caching strategies. Critical semantic data (e.g., detected anomalies) is stored in non-volatile memory with prioritized retention, whereas transient raw data is cached temporarily in overwritable buffers. This tiered approach guarantees persistence of essential information through communication blackouts. By systematically integrating acquisition, feature extraction, filtering, and storage, the Data Layer delivers refined, task ready semantic inputs to the Transport Layer while operating within stringent resource constraints of space environments.

\subsection{Transport Layer}

The Transport Layer serves as the semantic-aware communication core, ensuring efficient and reliable information exchange across space networks through task driven encoding, adaptive transmission, and semantic robustness mechanisms.

Semantic Encoding employs neural architectures to convert multimodal data into task-oriented compact representations. An embedding layer maps raw inputs into a unified semantic space, capturing high-level features, while a transformer encoder extracts cross-modal correlations through self-attention mechanisms. A dense layer filters mission-critical features, suppressing noise and amplifying essential patterns. These compressed representations become the foundation for adaptive transmission, where the layer dynamically adjusts encoding granularity (e.g., reducing transformer layers during signal attenuation) and prioritizes data streams based on a triage of factors: real-time channel quality, satellite positional relationships, and mission urgency.

Building on this adaptive framework, Semantic Decoding reconstructs information through context-aware neural modules. A reshape layer restores compressed vectors into structured formats, while a transformer decoder fills missing regions via positional encoding. A softmax layer resolves ambiguities through probabilistic outputs. Crucially, this decoding phase integrates semantic error correction, where domain knowledge validates consistency, ensuring fidelity even with partial data corruption.

Based on the aforementioned semantic encoding and decoding framework, the system further achieves real-time transmission adjustments by leveraging compressed semantic representations to dynamically refine encoding granularity according to channel quality, satellite positioning, and mission urgency, ensuring that high-priority data like collision warnings receive dedicated bandwidth while routine telemetry adopts a lighter mode. Simultaneously, context-aware error correction during decoding effectively identifies and replaces anomalies, such as implausible lunar atmospheric pressure readings, with context-derived defaults to maintain high semantic fidelity even with partial data loss, while embedded standardized metadata ensures that legacy ground stations can accurately parse next-generation probe data, preserving backward compatibility.

Through integrated task-driven encoding, adaptive transmission with granularity control and priority allocation, knowledge-guided error correction, and standardized metadata, the transport layer ensures reliable space network operations. This cohesive framework maintains semantic fidelity during bandwidth fluctuations, resolves data anomalies through domain constraints, and enables cross-generational system interoperability in challenging communication environments.

\subsection{Application Layer}

The Application Layer serves as the mission-centric intelligence core, driving a three-phase workflow, interpretation, decision, and evolution, to translate semantic information into context-aware decisions and sustained knowledge adaptation.

Semantic Interpretation initiates task execution by aligning incoming semantic data with domain-specific knowledge bases. Planetary surface imagery is analyzed against geological databases to identify landing hazards, while spectral telemetry cross-references mineralogical models to detect anomalies, enabling applications like Martian dust storm prediction or equipment health diagnostics through anomaly pattern recognition.

Building on these interpreted semantics, Contextual Decision Making dynamically generates actionable commands. Satellite swarms reroute formations based on proximity telemetry risks, while rovers adjust navigation paths for terrain obstacles. Concurrently, emergency protocols prioritize resource allocation, such as redirecting orbital assets to monitor volcanic eruptions, using real-time threat severity assessments.

To close the optimization loop, Knowledge Base Update refines repositories through feedback-driven adaptation. Mission metrics (e.g., asteroid detection false alarms) and operational data (e.g., model inference accuracy) are analyzed to update both local and global knowledge bases. This tri-phase cycle, interpretation to contextualize data, decision-making to trigger actions, and knowledge evolution to optimize future responses, ensures autonomous adaptation to dynamic space environments while maintaining alignment with long-term exploration objectives.

\subsection{Standardization and Interoperability Considerations}

The deployment of a semantic-enabled Internet of Space (IoS) requires robust standardization and seamless interoperability across heterogeneous space assets. To achieve this, the following key aspects must be addressed:

\paragraph{Unified Semantic Message Format}  
A key step is to define a CCSDS/ITU-aligned semantic packet format, co-developed with industry stakeholders, which encapsulates domain-specific semantic tags (e.g., telemetry, alert, scientific-observation), task identifiers, and adaptive metadata fields \cite{zhang2024unified}. This format must adhere to IEEE-ISTO interoperability guidelines to ensure semantic consistency across heterogeneous assets, while allowing vendor-specific extensions through reserved fields to prevent fragmentation

\paragraph{Compatibility with Existing Standards}  
The architecture integrates a two-way translation layer, endorsed by CCSDS and IEEE working groups, to bidirectionally convert legacy data (e.g., CCSDS TM/TC frames) into semantic packets. This layer embeds versioning control and fallback mechanisms, enabling backward compatibility with legacy satellites while providing a migration path for phased adoption. Compliance with ITU’s semantic interoperability framework (under development) is prioritized to align with global deployment roadmaps.

\paragraph{Interoperability and Future Expansion}  
The modular architecture implements standardized plug-in interfaces (per ITU-T G.semIoT recommendations) to integrate emerging technologies, such as AI-enhanced sensors or quantum relays. Each module’s metadata schema follows ISO/IEC 21838-compliant ontology templates, ensuring interoperability across vendors. To prevent divergent implementations, a mandatory semantic conformance certification process, jointly administered by CCSDS and IEEE, is enforced for all third-party extensions.



\begin{figure}
    \centering
    \includegraphics[width=1\linewidth]{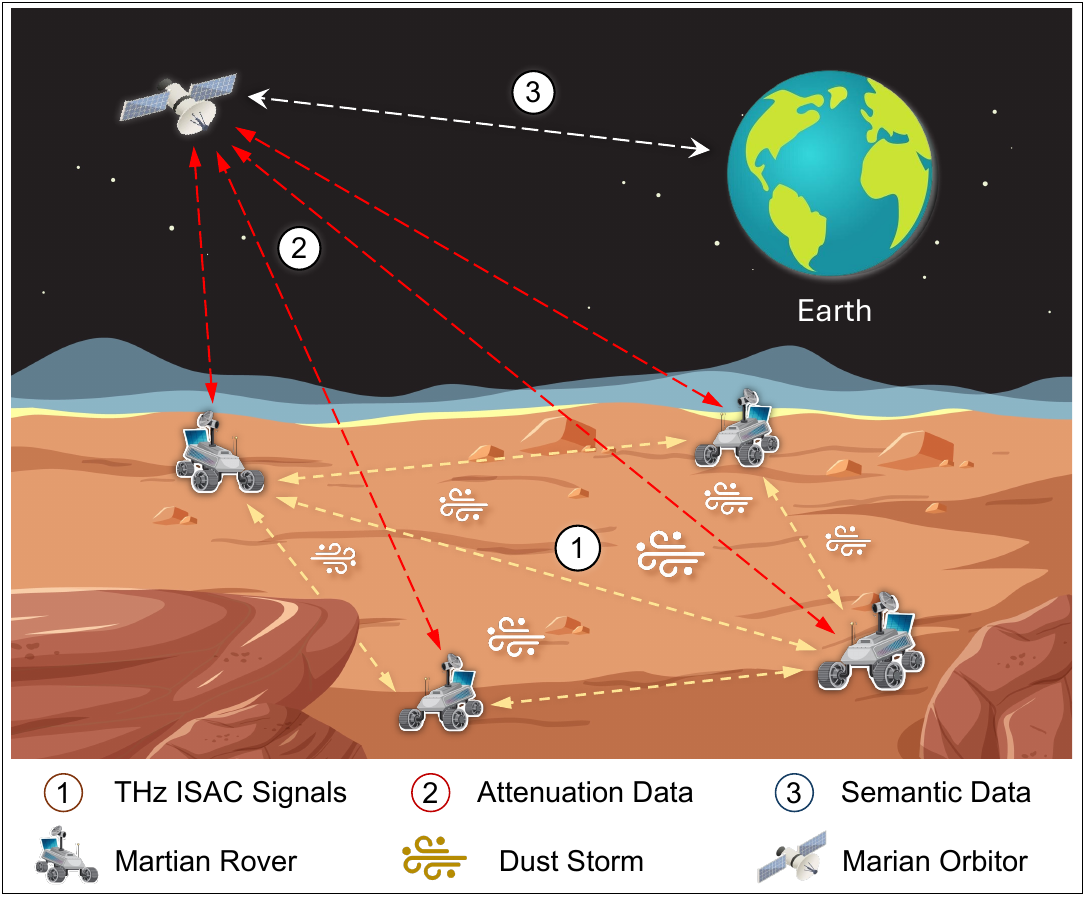}
    \caption{Illustration of semantic-based Mars surface exploration under dust storm conditions.}
    \label{fig:fig4}
\end{figure}

\section{User Case Scenarios}

To demonstrate the practicality and effectiveness of the proposed semantic communication architecture, we present a representative deep-space scenario focusing on semantic-based monitoring of Mars dust storms, as illustrated in Fig.~\ref{fig:fig4}. This scenario highlights the capability of the architecture to reliably detect, interpret, and communicate mission-critical events under extreme latency, limited bandwidth, and severe environmental interference conditions characteristic of interplanetary exploration. Such practical application underscores the significant advantages offered by semantic-enabled IoS architectures in addressing realistic operational challenges for future space missions.

In this deep-space scenario, Mars surface devices, such as rovers and landers, monitor environmental conditions during dust storms that can disrupt operations by obscuring solar panels, limiting mobility, and interfering with radio signals \cite{zhang2024analysis}. Real-time data collection and transmission under these harsh conditions are challenging. To address this, we implement the proposed semantic communication architecture through its three-layer framework, combining ISAC with standardized protocols to enhance efficiency, reduce resource demands, and ensure backward compatibility. A key innovation lies in repurposing unavoidable communication signal distortions, such as signal attenuation and scattering caused by dust particles, as indirect environmental metrics, thereby enabling effective real-time storm characterization and monitoring without imposing additional resource burdens.

\begin{figure*} 
    \centering 
    \includegraphics[width=1\linewidth]{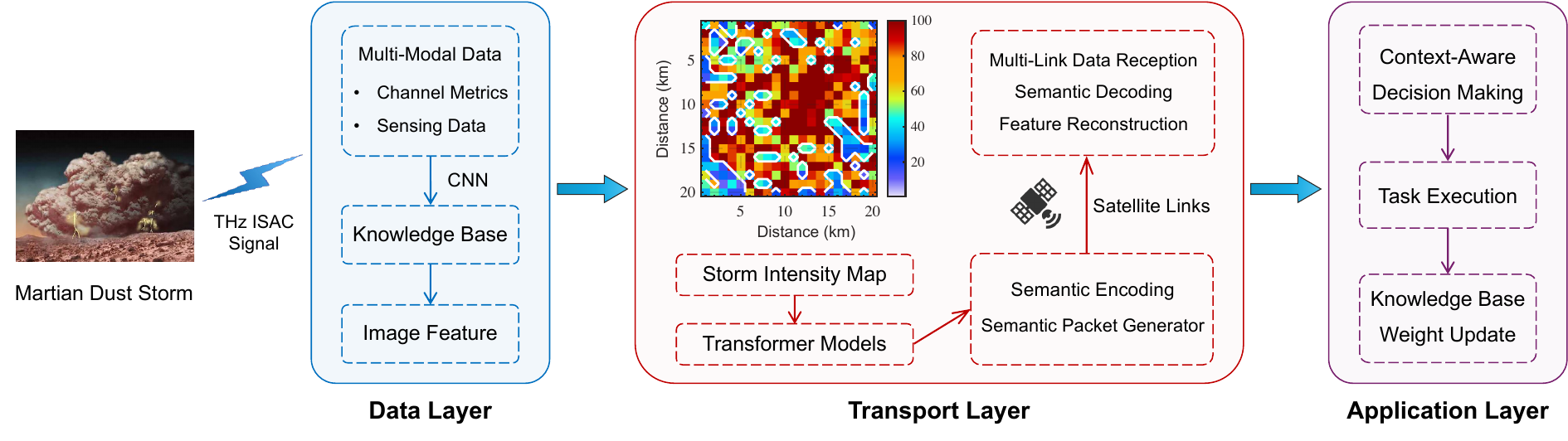}
    \caption{Illustration of the semantic communication pipeline for Mars dust storm monitoring based on the proposed IoS architecture.
    }
    
    \vspace{-10pt}
    \label{fig:data_representation}     
\end{figure*}


As illustrated in Fig.~\ref{fig:data_representation}, the Data Layer leverages THz-enabled ISAC to collect dust storm data in real time through opportunistic signals. As Mars devices communicate with each other and with orbiting relay satellites, their radio signals interact with dust particles—experiencing attenuation, scattering, and propagation delays. Rather than treating these distortions as mere interference, the system repurposes them as indirect storm metrics (e.g., dust density along communication paths). These metrics are then combined with direct sensor measurements such as wind speed, particle concentration, and visibility through signal inversion techniques and a predefined attenuation-to-concentration model. This approach enables real-time storm severity estimation without requiring additional hardware. A lightweight CNN, optimized for edge deployment on Mars devices, subsequently fuses these multimodal inputs to extract high-level semantic features. These features are systematically encoded into a knowledge base, which dynamically aggregates spatiotemporal storm patterns across missions through attention-weighted semantic embeddings. By processing data locally, the system minimizes the need to transmit large volumes of raw data, thereby conserving bandwidth and energy for critical operations.

After local extraction of semantic features, the Transport Layer packages them into compact, mission-aware representations, which a Transformer-based model then converts into efficient semantic packets. These packets are transmitted from Mars devices to Earth via relay satellites using a CCSDS-compliant format, ensuring compatibility with existing communication infrastructures. This efficient encoding optimizes bandwidth usage and facilitates integration with legacy ground stations while supporting the gradual adoption of semantic communication techniques.

At Earth-based control centers, the Application Layer decodes the semantic packets using context-aware neural modules and domain-specific knowledge bases tailored to Mars' unique conditions. This decoding process reconstructs the transmitted features into actionable insights, such as identifying storm patterns and assessing their impact on mission operations. For example, alerting controllers that a dust storm is obstructing a planned rover path. Continuous updates to the knowledge bases further refine the decoding process, supporting adaptive and real-time decision-making.

This semantic communication framework offers transformative benefits for deep-space missions by significantly reducing data transmission volumes while maintaining high accuracy. It addresses the bandwidth, energy, and latency challenges of Mars exploration. The dual use of communication signals for both data transmission and environmental sensing reduces the need for additional sensors, thereby lowering power consumption. Furthermore, the CCSDS-compliant packet format ensures smooth integration with both modern and legacy ground stations. Overall, this approach enhances system resilience, operational efficiency, and scientific discovery in the extreme environment of Mars.

\begin{figure}[t] 
    \centering 
    \includegraphics[width=1\linewidth]{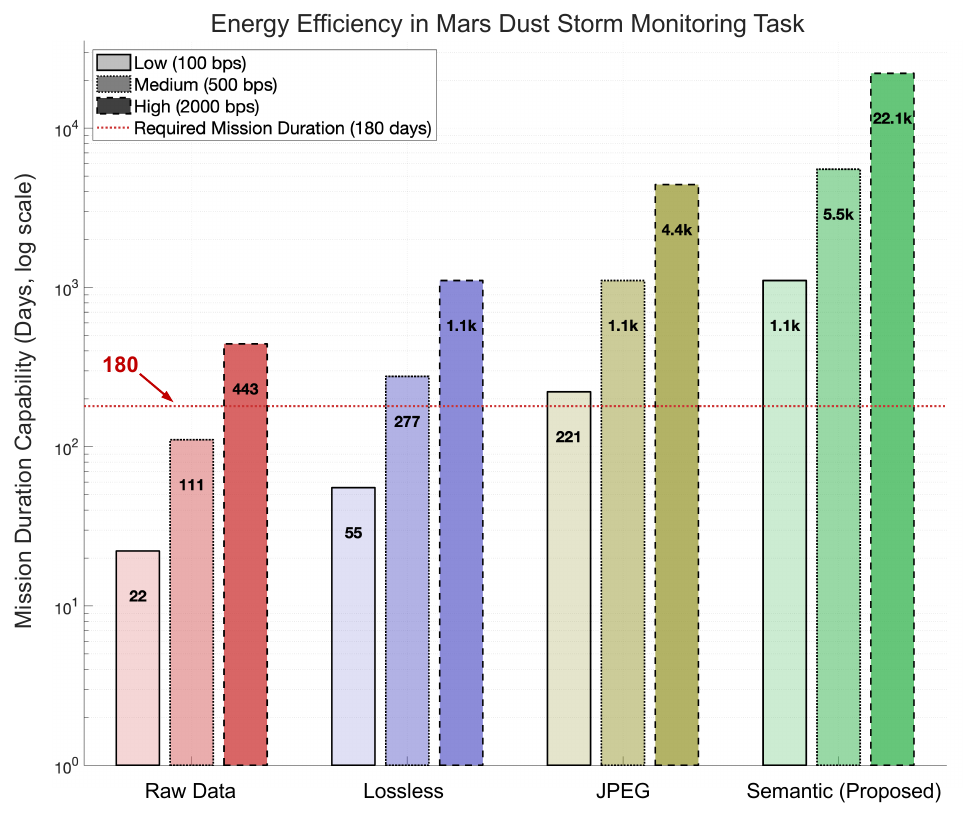}
    
    \caption{
    Energy efficiency comparison for Mars dust storm monitoring task.
    }
    \vspace{-12pt}
    \label{fig:energy_efficiency} 
\end{figure}

Fig.~\ref{fig:energy_efficiency} quantitatively evaluates the energy efficiency benefits of the proposed semantic communication architecture in the Mars dust storm monitoring scenario. Compared to conventional raw data transmission, the semantic approach significantly reduces transmitted data volume, thereby enabling up to 50 times more transmissions per battery charge. Notably, while conventional methods fail to meet the required 180-day mission duration at lower data rates (e.g., 22 days at 100 bps and 111 days at 500 bps), the semantic communication architecture consistently surpasses this threshold across all data rates. This directly addresses the severe energy constraints and operational reliability demands of Mars surface exploration, ensuring long-term mission sustainability.

Beyond the presented Mars scenario, our proposed semantic architecture can be broadly applied to various other challenging space exploration missions. Potential applications include semantic-driven asteroid characterization, where critical surface features and hazards can be efficiently encoded and transmitted back to Earth, and lunar exploration missions, enabling real-time anomaly detection in complex lunar surface operations. Additionally, the architecture is well-suited for deep-space observatories, facilitating semantic extraction and transmission of astronomical event data, such as transient phenomena. These examples further underscore the versatility and potential of our semantic-enabled architecture in addressing diverse operational challenges within future IoS missions.





\section{Open Research Challenges}

Semantic communication for the Internet of Space (IoS) presents numerous promising opportunities; however, significant research challenges remain unsolved to ensure practical feasibility and widespread adoption. In this section, we outline key open issues requiring immediate attention from academia, industry, and standardization bodies to fully realize the potential of semantic-enabled IoS.

\subsection{Efficient Multi-Modal Semantic Fusion with Physical-Layer Integration}

IoS networks face critical challenges in fusing heterogeneous data streams (e.g., telemetry, imagery, radar) under stringent resource constraints. Integrated sensing and communication (ISAC) paradigms offer a transformative approach by designing dual-functional waveforms that concurrently optimize semantic feature extraction and data transmission. For planetary exploration missions, such frameworks could unify high-resolution environmental sensing (e.g., terrain mapping via spectral analysis) and reliable communication through shared waveform architectures. Reconfigurable intelligent surfaces (RIS) further enhance adaptability by dynamically adjusting channel propagation characteristics to prioritize mission-critical modalities, such as reinforcing LiDAR data integrity during electromagnetic interference events.

The convergence of federated learning and physical-layer innovations introduces novel scalability challenges. Satellite constellations leveraging RIS-assisted beamforming could theoretically synchronize semantic models while performing distributed sensing tasks \cite{liao2021blockchain}. However, key open problems persist: 1) Joint waveform-semantic optimization balancing feature resolution with interplanetary channel degradation, 2) RIS-enabled attention mechanisms for real-time modality weighting under intermittent connectivity, and 3) Robust federated aggregation strategies resilient to dynamic Doppler shifts and propagation delays. Addressing these requires co-designing lightweight neural architectures with THz-band ISAC and RIS control protocols, ultimately enabling context-aware fusion that autonomously adapts to evolving mission requirements and harsh space environments.

\subsection{Dynamic Adaptation and Robustness}

The IoS network environment is highly dynamic, characterized by intermittent satellite connectivity, unpredictable solar interference, and significant propagation delays, such as the minutes-to-nearly-an-hour delays seen in Earth-to-Mars links. Traditional semantic encoding methods, typically designed for stable terrestrial channels, fall short under these conditions. Based on the aforementioned semantic encoding and decoding framework, there is an open research challenge to develop adaptive encoding methods that leverage compressed semantic representations to dynamically refine encoding granularity according to real-time channel quality, satellite positioning, and mission urgency. This approach ensures that high-priority data, such as collision warnings, receive dedicated bandwidth while routine telemetry adopts a lighter mode.

At the same time, semantic transmission is susceptible to errors from cosmic radiation, hardware malfunctions, or adversarial interference, which can corrupt critical packets like spacecraft health data or remote sensing alerts \cite{kodheli2020satellite}. To address this vulnerability, research must focus on robust semantic inference methods that incorporate context-aware error correction, effectively identifying and replacing anomalies (for example, implausible lunar atmospheric pressure readings are substituted with context-derived defaults) to maintain high semantic fidelity even amid partial data loss. Furthermore, embedding standardized metadata ensures legacy ground stations can accurately parse next-generation probe data, preserving backward compatibility.


\subsection{Standardization Framework and Evaluation Metrics}

The successful deployment of semantic communication in the Internet of Space (IoS) hinges on the establishment of standardized evaluation metrics that ensure interoperability, reliability, and mission effectiveness across diverse space assets. While existing communication standards from organizations such as CCSDS, IEEE, and ITU provide guidelines for syntactic-level data transmission, they fail to address key aspects of semantic communication, including meaning preservation, contextual accuracy, and the ability to support critical decision-making in space missions—factors essential for effective information exchange in space environments.

Traditional metrics such as bit error rate (BER) and packet delivery ratio (PDR) primarily assess transmission fidelity but fail to capture the semantic relevance and operational impact of received information. A standardized evaluation framework must therefore extend beyond conventional performance indicators to incorporate metrics that quantify semantic fidelity, context awareness, and task-specific accuracy. This includes assessing the degree to which transmitted information retains its intended meaning, evaluating its alignment with task objectives and environmental conditions, and measuring the system’s effectiveness in detecting anomalies with minimal false positives. Furthermore, the impact of semantic communication on decision-making processes in space operations must be systematically quantified to ensure that it enhances overall task performance. To establish a unified benchmarking methodology, collaboration among standardization bodies, research institutions, and industry stakeholders is essential. This includes defining reference models, developing testbed validation procedures, and establishing compliance criteria that facilitate performance comparison across different methods.

\section{Conclusion}

This article has presented a comprehensive vision for semantic communication in the 6G Internet of Space (IoS), outlining a novel architecture designed to meet the stringent demands of space-based communication environments. By shifting the communication paradigm from conventional bit-level transmission to meaning-driven semantic exchange, the proposed framework significantly enhances network efficiency, reliability, and adaptability in handling multi-modal data under challenging conditions. Through illustrative deep-space scenario, we demonstrated the practicality and effectiveness of our semantic-enabled IoS architecture. Despite promising advancements, several critical challenges, including efficient multi-modal semantic fusion, adaptive real-time semantic encoding, standardization, and robustness to semantic errors, require further research. Addressing these issues through collaborative research will be pivotal in transforming IoS from passive relay infrastructures into intelligent, context-aware communication networks capable of supporting future space missions and global connectivity objectives.




\ifCLASSOPTIONcaptionsoff
  \newpage
\fi



%


\bibliographystyle{IEEEtran}

\end{document}